\documentclass{aip-cp}

\usepackage{etoolbox}
\makeatletter
\patchcmd{\@@tablenote}{\xdef}{\protected@xdef}{}{}
\makeatother
\usepackage{etoolbox}
\makeatletter
\patchcmd{\@@tablenote}{\xdef}{\protected@xdef}{}{}
\makeatother
\usepackage[bottom]{footmisc}
\usepackage[numbers]{natbib}
\usepackage{rotating}
\usepackage{graphicx}
\usepackage[T1]{fontenc}
\usepackage{subfig}

\begin{document}

\title{Schwinger Generating Functional Derivation of LHC Elastic Proton-Proton Scattering Amplitudes}

\author[aff1]{Peter H. Tsang\corref{cor1}}

\affil[aff1]{Brown University, Providence, RI 02912. USA}

\corresp[cor1]{email: peter\_tsang@brown.edu}

\maketitle

\begin{abstract}
A recent Schwinger Generating Functional based formulation, by H.M.Fried,Y.Gabellini,T.Grandou and Y-M.Sheu and P.H.Tsang provided gauge-invariant, exact, non-perturbative solutions for QCD. After choosing a renormalization scheme and assuming a simpler 2 body quark-quark scattering problem, this formalism is tested against experimental elastic proton-proton scattering at LHC energies (7, 8, 13 TeV). This formulation was previously compared with ISR energies at (23.5 - 62.5 GeV). The full scattering amplitude with infinite gluons summed, infinite loops summed and their interference terms are needed for LHC differential cross-sections.
\end{abstract}
\footnote{Speaker. 16th International Conference of Numerical Analysis and Applied Mathematics : Analysis of Quantum Field Theory IV, Rhodes, Greece, 13-18 Sept, 2018.}
\section{INTRODUCTION}

\section{Schwinger's Generating Functional}

\begin{figure}[h]
  \centerline{\includegraphics[width=250pt]{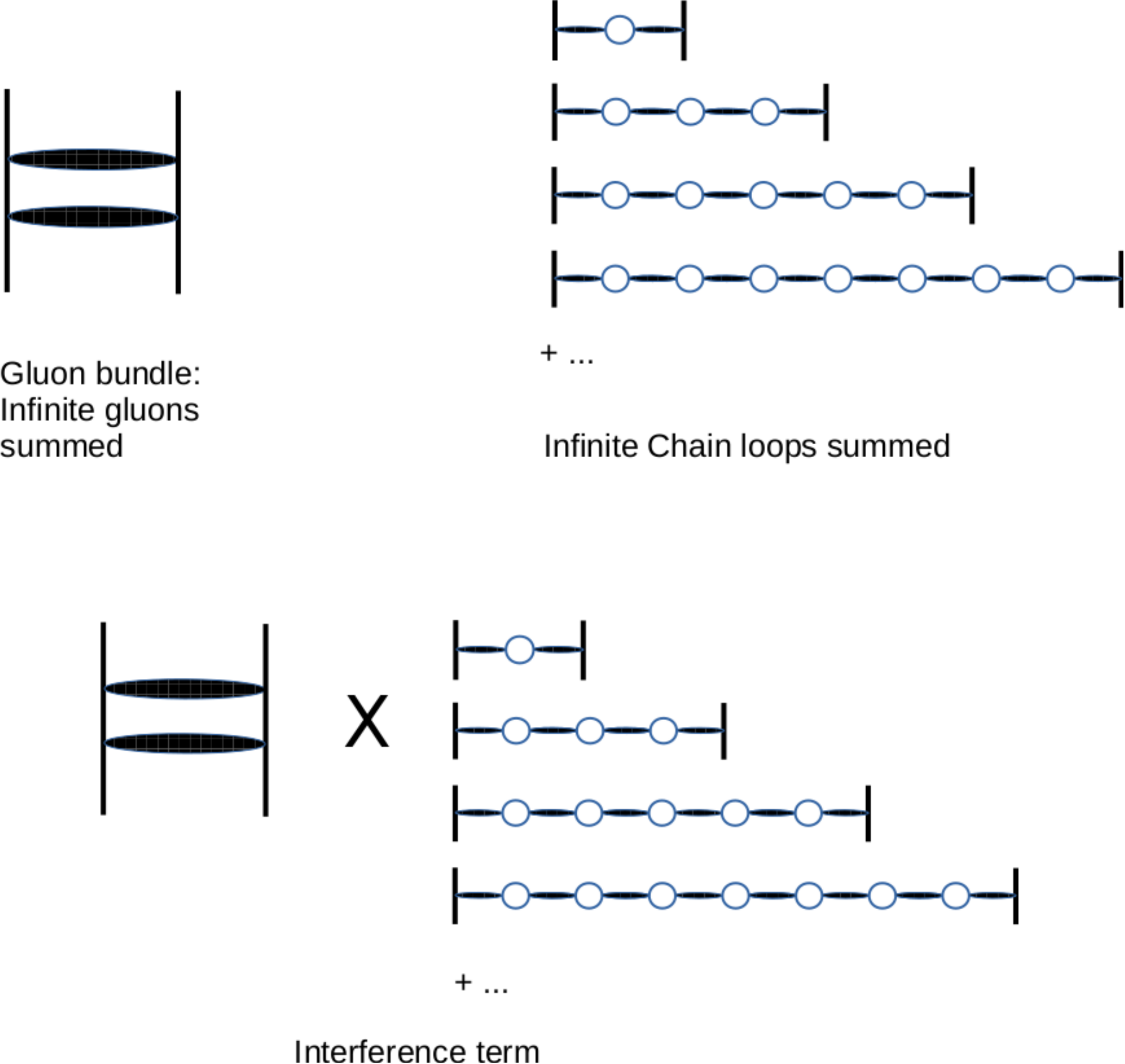}}
\caption{Interference term between Gluon Bundles and Infinite sum of chain loop diagrams.}
\label{interference}       
\end{figure}

\begin{figure}[!tbp]
\centering
\includegraphics[width=12cm]{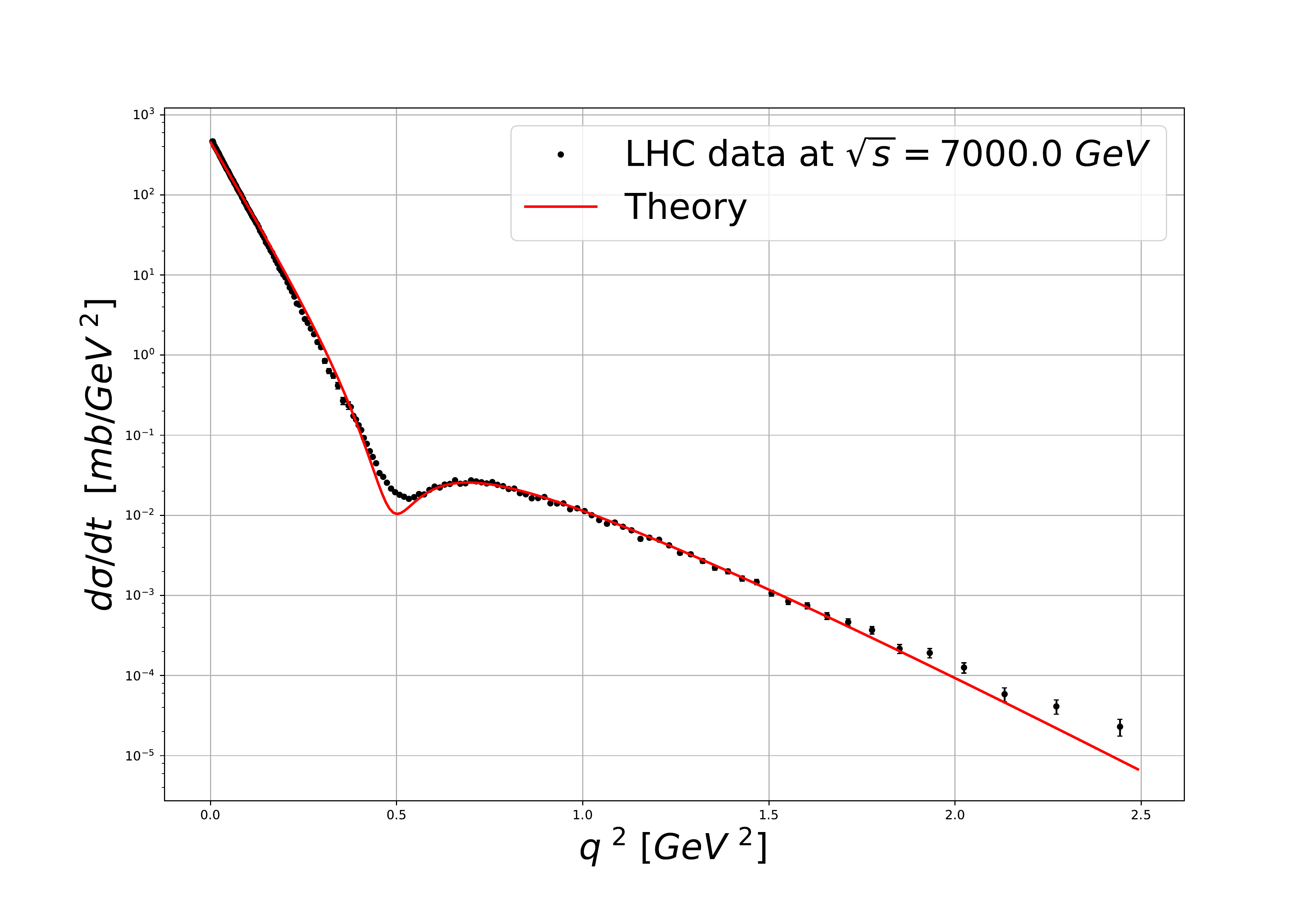}
\caption{Elastic pp scattering differential cross section at $\sqrt{s} = 7$ TeV. Black dots are experimental data, red line is the result from Eq.(\ref{differentialcrosssection})}
\label{7tev}       
\end{figure}

\begin{figure}[!tbp]
\centering
\includegraphics[width=12cm]{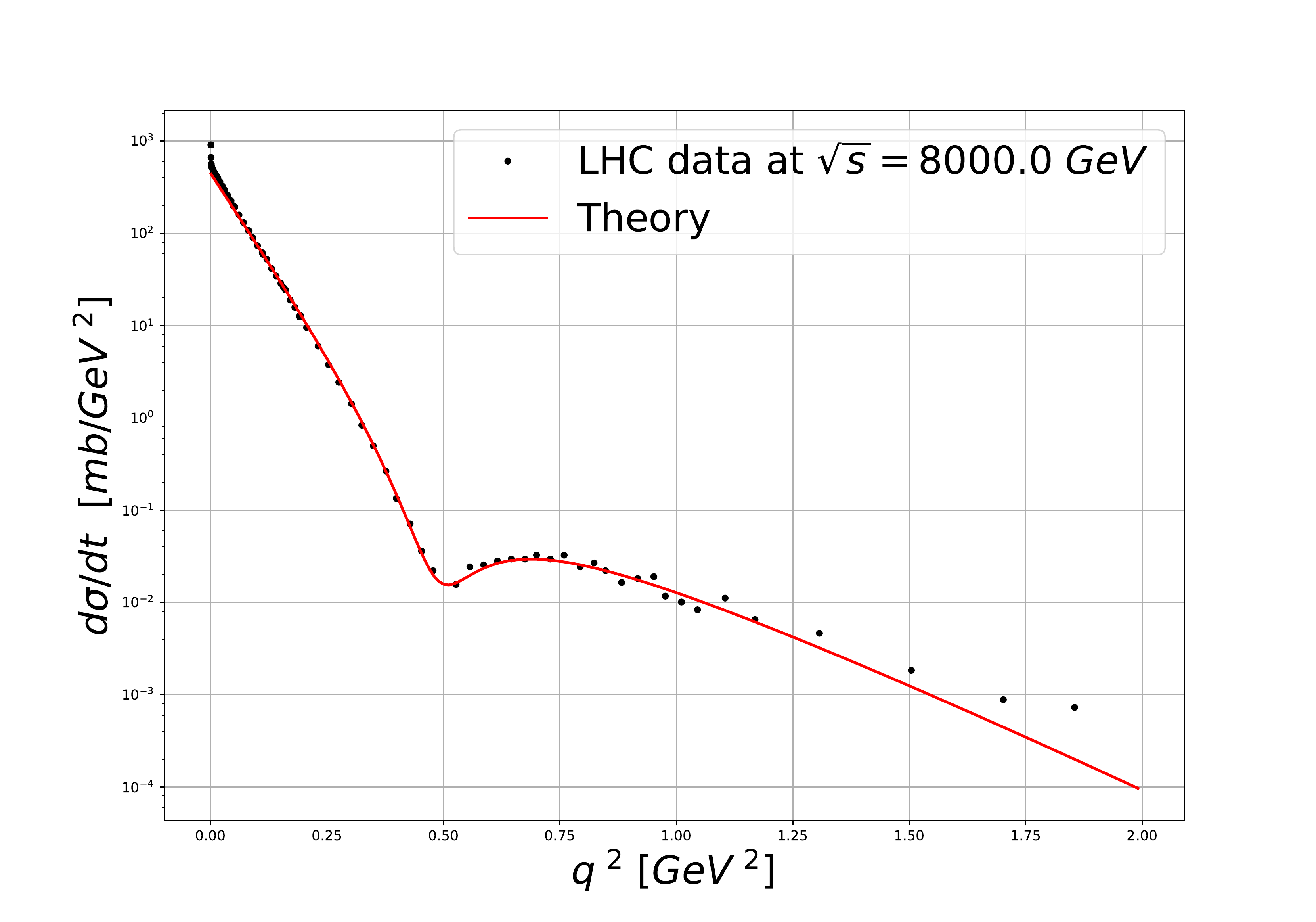}
\caption{Elastic pp scattering differential cross section at $\sqrt{s} = 8$ TeV. Black dots are experimental data, red line is the result from Eq.(\ref{differentialcrosssection})}
\label{8tev}       
\end{figure}

\begin{figure}[!tbp]
\centering
\includegraphics[width=12cm]{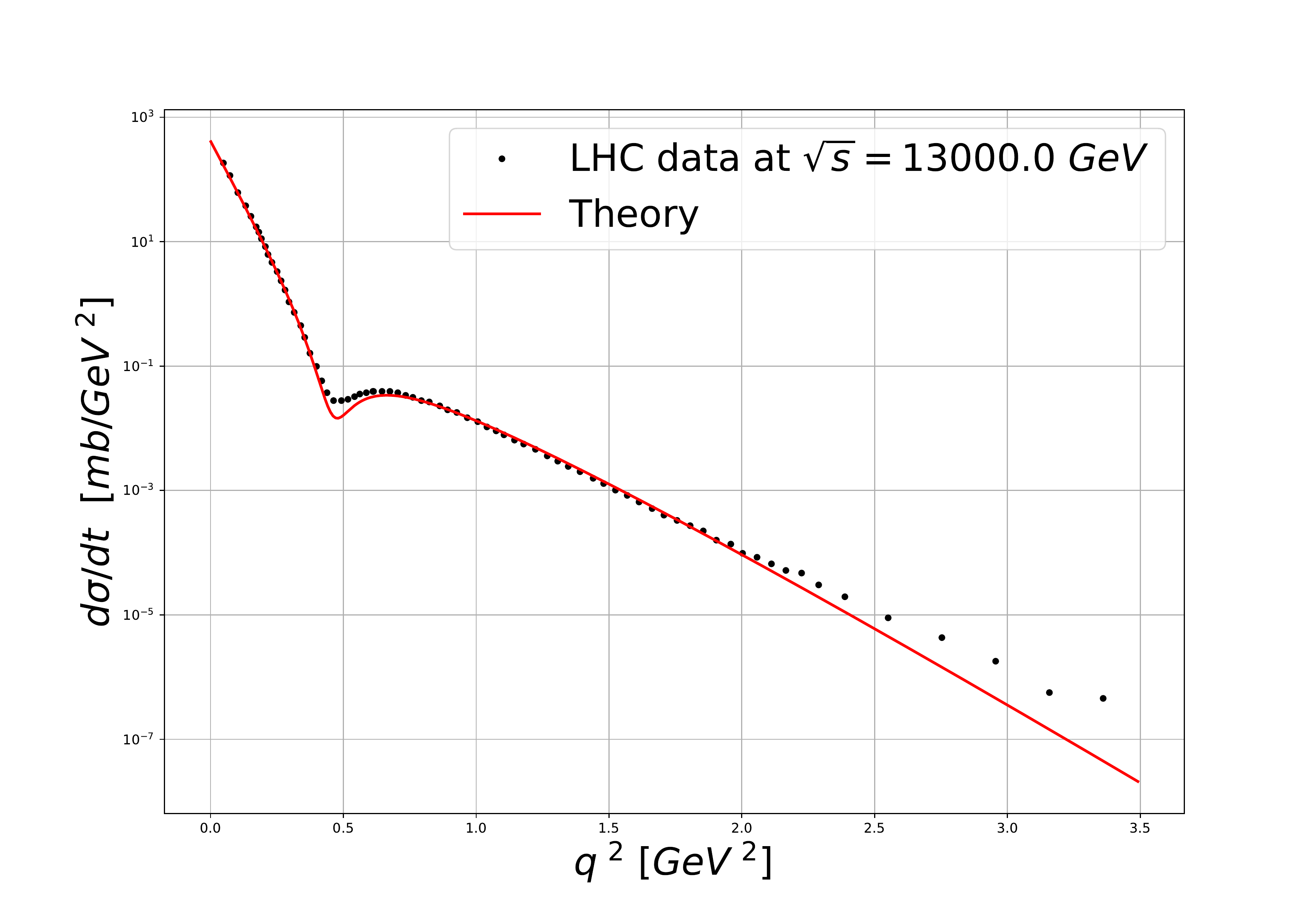}
\caption{Elastic pp scattering differential cross section at $\sqrt{s} = 13$ TeV. Black dots are experimental data, red line is the result from Eq.(\ref{differentialcrosssection})}
\label{13tev}       
\end{figure}

A recent Schwinger Generating Functional derivation of QCD correlation functions provided the total sum of all gluonic exchanges between two quarks\cite{qcd1}-\cite{qcd8}. After renormalization \cite{qcd8}, this formulation calculated QCD scattering amplitudes that included all gluonic exchanges, all loops to all orders. This was previously compared with proton-proton elastic scattering differential cross-sections at the ISR energies\cite{isr,peterthesis}. However, the interference terms between infinite sum of gluonic exchanges (Gluon Bundles), and the infinite loops exchanges were not included. It is shown in this talk that the interference terms, Figure~\ref{interference}, are essential for the shape of the diffraction dip in all elastic pp-scattering experiments, in particular at the LHC energies (7 TeV, 8 TeV, 13 TeV)\cite{totem}-\cite{totem3}, Figure~\ref{7tev}-Figure~\ref{13tev}.

\subsection{Gauge-Invariance}

Gauge-Invariance is insured by means of explicit Gauge Independence as shown in previous papers\cite{gaugeinvariance}.

\subsection{LHC data for elastic proton-proton scattering}

The scattering amplitude is derived from the Eikonal limit, where $\vec q^{\,2} = |t |<< s$, where the center of mass energy $s$ is much larger than momentum exchanged, $t$. The amplitude is then 

\begin{eqnarray}
T(s,\vec q)& =\frac{is}{2m^{2}}\int d^{2}b\ e^{ i\vec q\cdot \vec b}\ [1-e^{ i{\bf X}(s,\vec b)}]
\end{eqnarray}.

 The exact eikonal term is 
 \begin{eqnarray}
e^{\displaystyle i{\bf X}(s,\vec b)} = N \int d[\chi] e^{\displaystyle i/4 \int \chi^2}[\det(f\!\cdot\!\chi)^{-1}]^{1/2}\ \mathcal{F}\big(k'(f\!\cdot\!\chi)^{-1}+iC(f\!\cdot\!\chi)^{-2}\big)
\end{eqnarray}
derived in Reference~\cite{qcd8} and \cite{isr}, where $\chi$ is the Halpern field derived in \cite{qcd1}-\cite{qcd8}. $\mathcal{F}$ represents all gluonic exchanges between two quarks including infinite sum of all loops to all orders: 

\begin{eqnarray}
\mathcal{F}\big(k'(f\!\cdot\!\chi)^{-1}+iC(f\!\cdot\!\chi)^{-2} = \Big(1- (e^{ik' \varphi(b)R^{-1}+iC\varphi^2(b)R^{-2}})\Big),
\end{eqnarray}

where:
\begin{eqnarray}\label{k}
 k' =   (1+i)\ \sqrt[]{\beta/2}\ \delta^2_{quark} \ g, 
\quad     C  =    g^2 \beta\ \delta_{quark}^2 \kappa,
\quad \ \varphi(b)  =  e^{-m^2 b^2}.
\quad R^2 = |(f\cdot \chi)^2|,\quad and 
\quad \ \delta_{quark} = (m/E)^p
\end{eqnarray}

The simplification of $R^2 = |(f\cdot \chi)^2|$ is discussed in detail in Reference \cite{isr}. The scattering amplitude is then 
\begin{eqnarray}
T(s,\vec q) =\frac{is}{2m^{2}}\int d^{2}b\ e^{\displaystyle i\vec q\cdot \vec b}\ \Big[1-\Big(1- (e^{ik' \varphi(b)R^{-1}+iC\varphi^2(b)R^{-2}})\Big)\Big],
\end{eqnarray} 
where the $e^{ik' \varphi(b)}$ term represents all gluonic exchanges, the Gluon Bundle, between two quarks, and the $e^{iC \varphi(b)}$ represents infinite closed quark loops to all orders. The sum of the closed quark loops is a simple geometric sum as previously shown in Reference~\cite{isr}. The cross term, or interference term between Gluon Bundles and Closed quark loops was left out in previous ISR analysis. This interference term is included in this LHC analysis, and we see that, it is this interference term that determines the shape of the diffraction dip.

The resulting differential cross-section is then  given by:
\begin{eqnarray} \hfill\displaystyle\frac{d\sigma}{dt} & \displaystyle = \frac{m^4}{\pi s^2}\, | T |^2\hfill
 \end{eqnarray}
resulting in:

\begin{eqnarray}
    \frac{d \sigma} {dt}  &=& \ k\  \bigg(-(9\cdot 3 \cdot 4)(\frac{ g^2}{4\cdot4\pi})\Big(\frac{m_{ext}}{E}\Big)^{4p}Exp[\frac{-q^2}{8m_{ext}^2}] \nonumber\\ \nonumber && 
    +\quad  (9\cdot 3 \cdot 6)\frac{ g^2}{4}\Big(\frac{m_{ext}}{E}\Big)^{2p}\frac{q^2}{m_{ext}^2}Exp[\frac{-q^2}{4m_{ext}^2}](\frac{\kappa}{1+(\frac{g}{4} \kappa \frac{q^2}{m_{int}^2}Exp[\frac{-q^2}{4m_{int}^2}])^2})  \nonumber\\  && 
     +\quad \frac{4374}{16}  g^4 \kappa \Big(\frac{m_{ext}}{E}\Big)^{6p}(\frac{2m_{ext}}{9\pi m_{ext}^2})\sqrt{\frac{2\pi}{3}}(12m_{ext}^2+ q^2)Exp[\frac{-q^2}{12m_{ext}^2}]\frac{1}{1+(\frac{g}{4} \kappa \frac{q^2}{m_{int}^2}Exp[\frac{q^2}{4m_{int}^2}])^2}\bigg)^2
\label{differentialcrosssection}
\end{eqnarray},
where the first term is the Gluon Bundles, the second term is the geometric sum of infinite loops, the last term is the interference term between Gluon Bundles and Loops term.


The parameters are 1) bare coupling constant, $g=6.0$, 2) Gluon Bundle mass term, $m_{ext}=0.113 GeV$, 3) Closed Quark Loops mass term, $m_{int}=0.263 GeV$, 4) Renormalization parameter, $\kappa=6\times10^{-7} \sqrt{E}$, and  5) Energy dependence $p=0.02$. Note that $p$ can be computed, although difficult, in principle as shown in reference \cite{qcd7}.


\section{ACKNOWLEDGMENTS}
This work is partially supported by the Julian Schwinger Foundation.


\end{document}